# THE CHANDRA IRON-L X-RAY LINE SPECTRUM OF CAPELLA

Ehud Behar, Jean Cottam, and Steven M. Kahn

*Columbia Astrophysics Laboratory, Columbia University, New York, NY 10027-5247* [1]

An analysis of the iron L-shell emission in the publicly available spectrum of the Capella binary system, as obtained by the High Energy Transmission Grating Spectrometer on board the Chandra X-ray Observatory, is presented. The atomic-state model, based on the HULLAC code, is shown to be especially adequate for analyzing high-resolution x-ray spectra of this sort. Almost all of the spectral lines in the 10 – 18 Å wavelength range are identified. It is shown that, for the most part, these lines can be attributed to emission from L-shell iron ions in the Capella coronae. Possibilities for electron temperature diagnostics using line ratios of $Fe^{16+}$ are demonstrated. It is shown that the observed iron-L spectrum can be reproduced almost entirely by assuming a single electron temperature of $kT_e$= 600 eV. This temperature is consistent with both the measured fractional ion abundances of iron and with the temperature derived from ratios of $Fe^{16+}$ lines. A volume emission measure of $10^{53}$ $cm^{-3}$ is calculated for the iron L-shell emitting regions of the Capella coronae indicating a rather small volume of $10^{29}$ $cm^3$ for the emitting plasma if an electron density of $10^{12}$ $cm^{-3}$ is assumed.

---

[1] E-mail addresses: behar@astro.columbia.edu, jcottam@astro.columbia.edu, skahn@astro.columbia.edu,

# 1. INTRODUCTION

The Capella binary system consists of well-separated G1 III and G8 III stars with an orbital period of 104 days (Hummel at al. 1994). As one of the strongest coronal x-ray sources in the sky, Capella has been observed with all of the available x-ray telescopes dating back to the 1970's (Catura, Acton, & Johnson 1975). Past investigations, prior to the launch of the Chandra X-Ray Observatory, have been based upon limited-resolution broadband measurements. The present spectrum, which has been made publicly available in conjunction with the Chandra Emission Line Project (Brickhouse & Drake 1999), was obtained with the High Energy Transmission Grating (HETG) spectrometer on board Chandra (Markert et al. 1995) as part of the HETG calibration program. It features many resolved spectral lines, and provides a much more detailed insight into Capella's coronae than previous observations. Along with the Low Energy Transmission Grating (LETG) observation of Capella presented by Brinkman et al. (2000), these are the first observations of this quality from stellar coronae, apart from the well-studied solar corona. By comparing the observed spectrum with detailed calculated spectra at various plasma conditions, we attempt to provide a better understanding of the Capella coronae and of the atomic processes that are responsible for the strong x-ray line emission.

Similar to the x-ray spectrum of the Sun, the Capella spectrum is dominated by numerous iron L-shell emission features accompanied by a few K-shell lines from other elements, such as oxygen, neon, magnesium, and silicon. As in other iron-rich astrophysical plasmas, the strong lines emitted by the highly abundant Ne-like $Fe^{16+}$ ion are the most prominent lines in the Capella x-ray spectrum. These lines have been predicted to be promising candidates for temperature diagnostics in hot plasmas (Raymond & Smith 1986). The temperature dependence of the collisional excitation rates from the ground state $2p^6$ to the excited configurations $2p^53s$, $2p^53d$, and $2p^54d$ is expected to be particularly useful. The radiative decays following these excitations result in well-resolved x-ray lines around 17, 15, and 12 Å, respectively. The advantage of these lines as diagnostic tools is that they all pertain to a single ionization state and therefore the analysis of their intensities is generally independent of the ionization balance calculations, which strongly rely on total ionization and recombination rate coefficients that are probably of questionable accuracy.

Most of the early studies of Capella have concluded that its spectrum is comprised of two or more temperature components. The low- and high- temperature components have been interpreted as arising from smaller and larger coronal loops (Mewe et al. 1982; Vedder & Canizares 1983). Different plasma emission models have been employed to explain the observed spectrum and emission measure as a function of the two temperature components by best-fit methods. The ionic fractional abundances for the various elements were incorporated in these models from the literature (e.g., Arnaud & Raymond 1992). A few models also allowed the elemental abundances to vary from their solar values. The spread of temperatures found from the different missions and by the various methods is larger than can be explained by observational uncertainties. These temperatures range from 130 eV and 680 eV by Bauer & Bregman (1996) to 400 eV and 1900 eV by Lemen et al. (1989). The large discrepancies in temperature are not too surprising if one takes into account the variety of codes and the high number of free parameters in the fitting procedure as well as the different spectral regions observed and the possibility for time-varying emission.

Observations with the Extreme Ultraviolet Explorer (EUVE) by Dupree et al. (1993), Schrijver et al. (1995) and Brickhouse et al. (2000) have established that a continuous distribution of temperatures is present in Capella, featuring a sharp peak in the emission measure (EM) at about 550 eV and a drop by a full order of magnitude in the EM at temperatures of 100 eV higher or lower than that. Brickhouse et al. (2000) have, however, encountered difficulties fitting the x-ray spectra of Capella obtained with the Advanced Satellite for Cosmology and Astrophysics (ASCA) with the EUVE-derived models and found that a two-temperature model actually provides a better fit to the ASCA spectra. It was also found in that paper that the inclusion of Fe-L lines from very high Rydberg states dramatically improved their spectral fits.

In addition to the temperature, there have been previous measurements of the electron density in the Capella coronae, mostly by using ratios of EUV lines originating from iron L-shell ions. Linsky et al. (1998) obtained the high electron-density value of a few times $10^{12}$ cm$^{-3}$, which is not inconsistent with the $4\times10^{11}$ - $10^{13}$ cm$^{-3}$ estimation by Dupree et al. (1993). In the solar corona, such high densities have been detected only during flares (Feldman & Doschek 1978; Feldman, Doschek, & Kreplin 1982). There has not been direct evidence for intense flaring activity in Capella. Brinkman et al. (2000) have recently employed He-like line ratios of carbon, nitrogen, and oxygen to measure an electron density between $10^9$ and $10^{10}$ $cm^{-3}$ for Capella. However, these low-Z ions are most likely formed at different coronal regions than the iron L-shell ions. X-ray lines from iron L-shell ions do not provide robust density-sensitive ratios of strong lines for hot plasmas at densities below $10^{13}$ $cm^{-3}$. Only at higher densities do collisional excitation processes start playing an important role in altering the line intensities by depleting the upper levels of forbidden lines. In order to reproduce correctly the line intensities for electron densities above $10^{13}$ $cm^{-3}$, the pure coronal approximation may not be sufficient. Instead, a more comprehensive collisional-radiative model, which includes the collisional transitions between excited levels, is needed.

In the present work, we focus on the spectrum of the Capella binary system in the wavelength range of 10 - 18 Å, as detected by the HETG spectrometer on board the Chandra X-Ray Observatory in August 1999. The well-resolved spectrum exhibits many individual spectral lines that were not resolved in earlier observations. The spectrum is matched with detailed level-by-level calculations, using the Hebrew University Lawrence Livermore Atomic Code (HULLAC) by Bar-Shalom et al. (1998), for seven charge states of iron, $Fe^{15+}$ to $Fe^{21+}$, which allow us to construct theoretical spectra at different plasma conditions. The incomplete account of the numerous iron-L lines in plasma models have often been cited as the main reason for spectral fitting difficulties in the x-ray band. One of the main goals of this work is to show that the atomic calculations can accurately reproduce the entire iron-L spectrum of Capella in this wavelength range, including its numerous weak features. The high spectral resolution of HETG allows the dissection of the emitted spectrum without needing global fitting methods. Starting with the simplest model, we assume a single electron temperature and electron density for the entire ensemble of iron charge states. This highly simplifying assumption turns out to be fairly adequate within the observational uncertainties of the present spectrum. This cannot be too surprising as the EM distribution found from EUV observations exhibits a very narrow and sharp peak. The Capella coronae, obviously, cannot be completely isothermal. However, since the actual shape of the EM peak is not well constrained, the single temperature simplification is a reasonable modeling assumption. The theoretical spectra together with the theoretical line ratios are calculated by using a collisional-radiative model for each ionization state. By fitting the lines from each charge state separately, we avoid the inaccuracies involved in computations of ionic fractional abundances. Conversely, by fitting the theoretical spectra obtained for several charge states of iron to the multi-ion observed spectrum, the empirical fractional abundances of the iron ions can be derived and compared with theoretical results from the literature. In the last part of this paper, a volume emission measure is calculated for the iron-L emitting regions of the Capella coronae.

## 2. THE OBSERVED SPECTRUM

Capella was observed on August 28, 1999 for 14.8 ks as part of the HETG calibration program. The standard Chandra X-ray Center (CXC) processing produced event-reconstructed and aspect-corrected data for the publicly available Chandra archive in the form of ‘fits’ files. We have post-processed these files using custom software to extract the spectra. The CXC aspect correction algorithm had left a ghosted image (170 seconds), which was removed from the data set. A spatial filter 30-pixels wide was applied along the images of both the Medium Energy Grating (MEG) and the High Energy Grating (HEG) to separate the two spectra. For each of these data sets, the events were then plotted in dispersion versus pulse-height space to separate the spectral orders. A second filter, optimized interactively using a simple CCD model, selected all events associated with each order. The events were then plotted in a histogram with 5 mÅ bins. The background in this observation is negligible; therefore no background subtraction is necessary.

The wavelength calibration was performed by first locating the zero-order centroid and then converting dispersion in pixel number to dispersion in wavelength, using the HETG geometry. The Advanced CCD Imaging Spectrometer (ACIS) chip gaps had not been included in the CXC pipeline when this observation was processed, and so they had to be determined by simultaneously optimizing the wavelength residuals between the positive and negative spectral orders for both the MEG and HEG. After this calibration process the wavelength uncertainty is ± 2 mÅ.

In order to determine the flux, the histogrammed spectra were divided by the ground-calibration effective-area curves that were reported by Schulz et al. (1999). A comparison of the resulting positive and negative first order flux for the MEG grating shows an agreement of 20% (or better) from 7 to 18 Å. This confirms the shape of the effective area curves and hence the relative flux across this wavelength band for MEG. Since the same comparison for HEG shows somewhat larger discrepancies, for the flux analysis only the MEG spectrum is used. The positive and negative first orders of MEG, which include the vast majority of the photon counts, have been added to improve the statistical reliability. Henceforth, the sum of these two orders will be referred to as the Capella spectrum. This spectrum in the 10 - 18 Å wavelength range is represented by the upper trace in Figs. 1a and 1b.

## 3. LINE IDENTIFICATION AND FLUXES

In order to facilitate the identification of this rich line spectrum, calculations have been performed for the iron ions $Fe^{15+}$ through $Fe^{21+}$. All of the prominent lines in the 10 - 18 Å wavelength band of the Capella spectrum are identified, as well as most of the weak ones. The majority of the lines originate from iron L-shell ions, but a few oxygen, neon and nickel lines are also identified. In addition to the present HULLAC calculations, the most useful references for identifying the lines are the solar observation by McKenzie et al. (1980) and the comprehensive Electron Beam Ion Trap (EBIT) iron measurements by Brown et al. (1998, 2000). The theoretical spectrum, calculated for iron only, is plotted in Figs. 1a and 1b below the observed spectrum. The details of the calculations are discussed in the following sections. The ranges corresponding to the most important types of transitions for the various iron ions are indicated above the spectrum. The individual line identifications are presented in Table I for the most prominent lines in accordance with the peak labels (first column) in Figs. 1a and 1b. In cases where a spectral feature is identified as a blend of several lines, only the strongest lines are listed in the table. The second column in Table I shows the wavelengths in Ångtröms measured in the present observation ($\lambda_{observed}$). The third, fourth, and fifth columns, respectively, give the wavelengths previously measured in the Sun ($\lambda_{Sun}$), with EBIT ($\lambda_{EBIT}$), and the wavelengths presently calculated with the HULLAC code ($\lambda_{HULLAC}$). The following columns indicate the emitting ion and the upper and lower configurations (in the jj coupling scheme) of the radiative transition, including the total angular momentum quantum numbers of the upper and lower levels, $J_U$ and $J_L$. It should be stressed that the actual atomic levels involved in the radiative process may be composed of components from more than one configuration (i.e., configuration mixing). In those cases, the configuration given in Table I represents the configuration of the major component of the level.

The line fluxes are measured by fitting the lines with Gaussian profiles, which reproduce the observed line profiles very well. The last column in Table I gives the measured line fluxes and the next to last column gives the counts in the 5mÅ bin corresponding to the line peak. Some of the closely blended lines require fitting by two Gaussians simultaneously. These blends are indicated in the table by having the same label, but with different lower case indices (e.g., 4a and 4b). An example of a two Gaussian fit is given in Fig. 2 for peak 12. Only features that were reproduced in all of the four MEG and HEG positive and negative first order spectra were considered for Table I. Among these, only lines that could be fit with a Gaussian of width 15 – 28 mÅ FWHM (Full Width at Half Maximum) are included. The peaks with the widest profiles (FWHM > 25 mÅ) are associated with weak unaccounted for blends and are indicated in the table by (w) next to the flux value. As can be seen by the labels in Figs. 1a and 1b, Table I comprises all of the major spectral features, but excludes minor features that could be suspected as noise. From the level of agreement between the positive and negative first order fluxes of the MEG grating, the uncertainty in the line fluxes is estimated at ~20%. However, for the very weak lines where the peak count (next to last column) is less than 20, slightly larger errors may be present due to insufficient statistics.

It can be seen in Table I that for most of the strong lines, the presently observed wavelengths are in excellent agreement with those measured in the Sun and in EBIT. A few lines, which were not identified by McKenzie et al. (1980), are identified here. In general, the present x-ray line emission from Capella and the solar flare spectrum (McKenzie et al. 1980) are strikingly similar. The light curve of the present observation shows no obvious evidence for flaring activity. In the EBIT experiments, the numerous iron lines have been identified with the aid of HULLAC calculations. The accuracy of the HULLAC wavelengths is usually to within 10 - 20 mÅ as can be seen in Table I. More importantly, it is the calculated line *intensities* that allow us to predict which lines (among thousands) will be observed in the spectrum and how intense they will be. In the next section, we describe the atomic-state model as well as its application to the present Capella spectrum.

It is worth pointing out that some very weak lines arising from decays of high-*n* levels, *n* being the principal quantum number of the excited electron, can be detected in the Capella spectrum around 10 Å (Fig. 1a), the strongest of these arising from $Fe^{16+}$. These lines are also present in the calculated spectrum. Brickhouse et al. (2000) have claimed that the lines with n > 5 contribute a considerable amount of flux to the Capella spectrum measured by ASCA around 10 Å. Since the total flux from Capella in that region is extremely small, the accumulated flux of the many weak high-*n* lines may be important in the low-resolution measurements of ASCA. In the present high-resolution spectrum, however, these lines have an insignificant effect due to their extremely weak individual intensities. The strongest line in this region is clearly the $Ne^{9+}$ Lyβ line at 10.240 Å (labeled 1). The 2p-6d line of $Fe^{16+}$ is labeled 2 in Fig. 1a and in Table I and its measured line flux is ~ $10^{-4}$ $s^{-1}cm^{-2}$. The higher members of the 2p-*n*d line series, e.g., at 10.504 Å ($n$ = 7) and at 10.333 Å ($n$ = 8), are even weaker and can hardly be distinguished from the noise. Thus, they are not included in Table I.

## 4. THE ATOMIC-STATE KINETIC MODEL

Stellar coronae are characterized by relatively high temperatures and low densities, in which the most important atomic processes are electron impact excitations from the ground states and subsequent radiative decays, either directly to the ground states or via cascades. The hot stellar coronae are usually, but not always, optically thin. In the present work, we use a full collisional-radiative steady-state model, which includes all of the collisional excitations and de-excitations as well as all of the radiative decays between the levels of each ionization state, but neglects line self-absorption processes. The electron energy distribution is assumed to be Maxwellian, corresponding to an electron temperature $T_e$. The set of rate equations for the (time independent) density $n_i^{q+}$ of an ion with charge $q$+ in a level $i$ can be written as:

$$\frac{d}{dt} n_i^{q+} = n_e \sum_{j \neq i} n_j^{q+} Q_{ji}(T_e) + \sum_{j>i} n_j^{q+} A_{ji} - n_i^{q+} \left( n_e \sum_{j \neq i} Q_{ij}(T_e) + \sum_{j<i} A_{ij} \right) = 0 \qquad (1).$$

$A$ represents the Einstein rate coefficient for spontaneous radiative emission, and $Q(T_e)$ the rate coefficient for electron-impact excitation or de-excitation. The electron density is denoted by $n_e$. Level-by-level calculations have been carried out for each of the iron ions from Na-like $Fe^{15+}$ through B-like $Fe^{21+}$. For the various ions, high lying levels with $n$–values as high as 12, but not less than 6 are included in the model. The set of equations (1) are solved by normalizing the sum of all level populations to unity. In other words, the population of each level is calculated in relation to the total population of ions in its ionization state. In the high-temperature and low-density limit, where collisional transitions between excited levels are negligible, this model reduces to the standard coronal model, which is sufficient for most of the present work. The line intensities in number of photons emitted per unit time and per unit volume are subsequently given by:

$$I_{ji} = n_j A_{ji} \qquad (2).$$

The basic atomic quantities used in this work were obtained by means of the multi-configuration, relativistic HULLAC (Hebrew University Lawrence Livermore Atomic Code) computer package developed by Bar-Shalom et al. (1998). HULLAC includes the RELAC code for atomic energy levels (Klapisch et al. 1977) and for radiative decay rates as well as the CROSS code for collisional excitation rate coefficients (Bar-Shalom, Klapisch, & Oreg 1988). The intermediate-coupling level energies are calculated using the relativistic version of the parametric potential method by Klapisch (1971), including configuration mixing. The collisional rate coefficients are calculated in the distorted wave approximation, implementing the highly efficient factorization-interpolation method by Bar-Shalom et al. (1988). For $Fe^{16+}$, resonance contributions to the excitations to $2p^5 3l$ ($l$= s, p), i.e. dielectronic capture followed by autoionization to excited levels, are incorporated from Chen & Reed (1989). For the inner-shell excited levels of $Fe^{15+}$, dieclectronic capture from and autoionization to $Fe^{16+}$ are also included.

## 5. RESULTS OF THE MODEL

The individual spectrum for each of the ions $Fe^{15+}$ through $Fe^{21+}$ is obtained by a collisional-radiative model at a single electron temperature corresponding to 600 eV and at a single electron density (temperature and density diagnostics are discussed in Sec. 6). Each of these spectra is plotted separately at the bottom of Fig. 3. The isonuclear trends of the dominant 2p-3d transitions can be nicely seen in the figure shifting from about 15.0 Å for $Fe^{15+}$ and $Fe^{16+}$ to 11.8 Å for $Fe^{21+}$. Furthermore, the intensity of these transitions is shown to diminish with the ion charge as the number of electrons available for excitation in the 2p shell decreases. The theoretical spectral lines in Fig. 3 and throughout this paper are given a uniform Gaussian profile with 15 mÅ FWHM, slightly less than the observed line widths. The total iron spectrum is calculated by summing the contributions of all of the charge states. In this sum, each charge state is given a different weight corresponding to its ionic fractional abundance. The non-iron lines and blends are not included in the theoretical spectrum, but are indicated above the observed spectrum by the numbers 1 - 3, corresponding, respectively, to lines from Ne, Ni, and O. The total calculated spectrum is normalized to the 2p-3d $Fe^{16+}$ line at 15.264 Å. The iron fractional abundances are determined by scaling the relative line intensities of each ionization state to the observed intensities. The resulting fractions are indicated in the figure just above each individual-ion spectrum. The agreement between the Capella spectrum, given at the top of Fig. 3, and the total calculated spectrum, given just below it, can be seen to be very good. In particular, the discrepancy between the observed and calculated intensities usually does not exceed 20%, which is approximately the uncertainty of the measured line fluxes. Larger discrepancies occur for the 2p-3s lines of $Fe^{17+}$ around 16 Å. This is due to the resonance contributions to the excitations to $2p^4 3s$, which are not available in the literature and, thus, not included in our model, unlike the resonant excitations to $2p^5 3s$ for $Fe^{16+}$, which are included. The model can be even further improved by incorporating experimentally measured wavelengths (Brown et al. 1998, 2000) and collisional excitation rate coefficients (e.g. Gu et al. 1999), where available.

The fact that the Capella iron x-ray spectrum is accurately reproduced with a single temperature does not, of course, imply that the entire Capella coronae are isothermal. As the spatially and temporally integrated observation cannot be expected to reflect the temperature and density gradients that exist within the Capella coronae nor their time variability, the present result should be viewed as providing the most appropriate average temperature value for the iron-L emitting regions of the Capella coronae during the course of this observation. The adequacy of the single temperature model is consistent with the sharply peaking EM distribution found from the EUVE observations. EUV lines from ionic species ranging from $Fe^{8+}$ to $Fe^{23+}$, which are necessarily emitted from lower and higher temperature regions, have been detected by Brickhouse et al. (2000). In the present spectrum, the strongest x-ray lines of $Fe^{23+}$ at 10.663 Å and 11.178 Å are clearly absent and those of $Fe^{22+}$ and $Fe^{21+}$ are extremely weak (see peak 10 in Fig. 1a) demonstrating that the contribution to the EM from temperatures higher than 600 eV is very small. The present HETG spectrum cannot provide information regarding the charge states lower than $Fe^{15+}$, since these ions usually form in colder regions and do not emit x-rays. It should be further noted that Bauer & Bregman (1996) have successfully fitted a broadband (6 - 60 Å) spectrum of Capella measured by the ROSAT x-ray telescope, with a single temperature corresponding to 680 eV. The present analysis of the high resolution HETG spectrum, however, constitutes a more detailed demonstration of the fact that one temperature and one density are sufficient for reproducing almost the entire iron L-shell x-ray spectrum of Capella.

In the determination of the ionic fractional abundances, the strongest 2p-3d line is used for each charge state, except for $Fe^{16+}$ and $Fe^{17+}$, for which self-absorption may affect the observed line intensities. Instead, for these two ions, the other 2p-3d lines are used. Indeed, as can be seen from Fig. 3, the intensities of the strongest lines of $Fe^{16+}$ and $Fe^{17+}$, at 15.015 Å and 14.206 Å, are overestimated in the calculations. The intensity ratio of the $Fe^{16+}$ lines at 15.015 Å and at 15.264 Å has been a long-standing subject of numerous investigations. Brown et al. (1998) have measured this ratio in EBIT to be between 2.77 and 3.15 (depending on the electron beam energy), whereas the values from solar observations range from 1.5 to 2.4. The ratio measured in the present Capella spectrum is 2.4, fairly close to the measured value of 2.6 recently reported for Capella by Brinkman et al. (2000). The low observed ratio has most frequently been attributed to self-absorption of the 15.015 Å line (Waljeski et al. 1994). According to our calculations, another reason for the attenuation of this ratio may be the blending of the 15.264 Å $Fe^{16+}$ line with a strong $Fe^{15+}$ line that can contribute about 10% to the intensity at 15.264 Å.

## 6. PLASMA DIAGNOSTICS

### 6.1 Electron Temperature

A natural application of the present calculations is to employ the resulting line intensity ratios for plasma diagnostics. A useful summary of the potential of temperature diagnostics with $Fe^{16+}$ lines is given by Raymond & Smith (1986). In general, the relative intensities within each charge state of most of the iron lines in the currently studied wavelength range show only weak dependence on the electron temperature in the 200 eV to 1000 eV regime. This fact is demonstrated in Fig. 4, which gives the total calculated spectrum at the same density, but for three different temperatures, compared with the observed spectrum. All spectra are normalized to the 2p-3d line at 15.264 Å. In order to simplify the comparison, the fractional abundances best fit for 600 eV have been retained for the other temperatures as well. It can be seen that only minor differences occur between the various calculated spectra. As the temperature increases, the 2p-3d lines become slightly more intense with respect to the 2p-3s lines. However, since resonant contributions to the collisional excitation rates play a major role in populating the $2p^5 3s$ levels, and since this introduces an additional complication to the line-intensity modeling, we choose not to use these lines for diagnostics.

Perhaps the most easily observed temperature effect on the iron L-shell lines is the increasing intensity with temperature of lines originating from highly excited levels. At low temperatures, the rates for collisional excitation to the high, upper levels of these transitions are very small. Only at higher temperatures do these levels become significantly populated. A good example for this is the 2s-3p line at 13.829 Å and the 2p-4d lines at 12.266 Å and 12.131 Å (all pertaining to $Fe^{16+}$). A similar temperature effect is predicted for the 2p-5d $Fe^{16+}$ lines at 11.251 Å and 11.129 Å. These latter lines are too weak in the present observation limiting the accuracy that can be expected from them as temperature diagnostics.

Fig. 5 shows the calculated intensities for the 2s-3p line at 13.829 Å and for the 2p-4d line at 12.266 Å, with respect to the intensity of the 2p-3d line at 15.264 Å, as a function of temperature. The observed ratios have been increased by 10% to account for the blending of the 15.264Å line with $Fe^{15+}$, as suggested by our calculations (Sec. 5). These corrected ratios are represented in the figure by the horizontal lines. The ratios both yield a temperature of approximately 540 eV. The uncorrected ratios would indicate electron temperatures of 450 eV and 420 eV. It should be noted, also, that the uncertainty in the line fluxes is translated into fairly large error bars for the derived temperature. With the current uncertainty of 20% in the line fluxes, it is hard to constrain the temperature to better than 540 eV ± 50%. The charge state distribution, discussed below in Sec. 6.3, is much more sensitive to the electron temperature and therefore can potentially serve as a better temperature probe. However, the charge state distribution is much harder to calculate than the excitation rates needed for individual line ratios. Thus, its predictions are less reliable.

## 6.2 Electron Density

The possibilities for electron density diagnostics with the iron L-shell x-ray lines at high coronal temperatures are even scarcer than those for the electron temperature. We have employed a full collisional-radiative model for all of the iron ions $Fe^{15+}$ - $Fe^{21+}$ to obtain spectra calculated at the same temperature (600 eV) and ionization balance conditions, but for densities that differ by four orders of magnitude, from $10^9$ through $10^{13}$ $cm^{-3}$. The calculated spectra are all found to be essentially identical to the total spectrum displayed in Figs. 3 and 4 for 600 eV. The x-ray lines become useful for diagnostics only at higher densities where the collisional rates become comparable to the x-ray radiative rates. Dupree et al. (1993) and Linsky et al. (1998) have used for density diagnostics EUV lines ($\Delta n = 0$) that are emitted by the same iron ions investigated in the present x-ray analysis. The density sensitivity of the EUV lines starts at lower densities than that for the x-ray lines. The electron density obtained from those EUV observations is about $10^{12}$ $cm^{-3}$.

Brinkman et al. (2000) have recently employed the He-like (K-shell) C, N, and O lines in the Capella x-ray spectrum measured with the LETG spectrometer on board

Chandra to yield electron densities between $10^9$ and $10^{10}$ $cm^{-3}$. However, the He-like lines from these low-Z elements largely come from coronal regions that are significantly cooler (< 100 eV) than those from which the iron L-shell lines are emitted. Therefore, the density measured with the iron EUV line ratios is probably more adequate for describing the present iron L-shell plasma than the low-Z K-shell densities.

### 6.3 Fractional Ion Abundances

The fractional abundances of the iron ions, which are obtained here from the comparison between the calculated and the observed spectra for $Fe^{15+}$ - $Fe^{21+}$ are: 0.06, 0.34, 0.25, 0.13, 0.11, 0.06, and 0.05, respectively. The abundances of iron ions with $q <$ 15 and $q >$ 21 are negligible. In Fig. 6, the derived abundances are compared with the ionization balance calculations by Mazzotta et al. (1998) for temperatures of 430 eV, 610 eV, and 860 eV. The observationally deduced values for Capella in the figure (squares) are given error bars of 20% in order to account for the uncertainties in the observed and calculated line intensities. This is most likely an underestimation of the errors for the less abundant ions $Fe^{15+}$, $Fe^{20+}$, and $Fe^{21+}$, which feature only very weak lines. Note that the calculated $Fe^{15+}$ abundance at 860 eV, as well as the $Fe^{20+}$ and $Fe^{21+}$ abundances at 430 eV, are too low to be plotted in the scale of Fig. 6. It can be seen that the present values derived for Capella are in very good agreement with the calculated values for 610 eV, especially for the ions that dominate the spectrum, $Fe^{16+}$ through $Fe^{19+}$. The calculated value for $Fe^{16+}$ at 610 eV falls right on top of the measured data point and is hardly seen in the figure. The temperature of about 600 eV indicated by the comparison with the ionization balance calculations is also highly consistent with the value derived from the line ratios in Sec. 6.1. The remaining discrepancies for the abundances of $Fe^{15+}$ and $Fe^{21+}$ may be due to the fact that these ions are emitted from different temperature regions. Conversely, it is well known that the ionization balance calculations for open-L-shell Fe ions, such as $Fe^{20+}$ and $Fe^{21+}$ are questionable as a consequence of the (perhaps best available, but nevertheless) inadequate dielectronic recombination rates they incorporate. Thus, these highly suspect values of the theoretical fractional abundances may also contribute to the observed discrepancies.

The ionization balance is much more sensitive to the electron temperature than the line ratios within any given charge state. This is well illustrated in Fig. 6 by the agreement of the observed abundances with those calculated for 610 eV as opposed to the complete disagreement with the calculations for 430 eV and 860 eV. The fact that the empirically derived charge distribution agrees well with a single temperature distribution provides more supporting evidence for the EM distribution found from the EUVE observations, i.e., a very narrow EM distribution that sharply peaks around a temperature of about 550 eV. The most important provision regarding the exact temperature structure of the Capella coronae remains, however, the questionable reliability of the ionization balance calculations.

## 7. VOLUME EMISSION MEASURE

The results of our coronal models for the various iron ionization states together with the measured flux in the present observation can be exploited to obtain a rough estimation of the volume emission measure *VEM*, and based on the electron density measured in previous papers, the volume *V* of the emitting plasma. The *VEM* is obtained from the measured photon flux *f*, i.e., the number of photons detected per second and per $cm^2$, and from the calculated number emissivity *F*, i.e., the number of photons emitted in the plasma per second and per $cm^3$, through the relationship:

$$VEM = n_e^2 V = n_e^2 \frac{4\pi d^2 f}{F} \qquad (3).$$

This form of the *VEM* assumes that there is no x-ray absorption and that all of the x-rays are emitted outward from the stars. The distance to Capella is denoted by *d*. The number emissivity *F*, which is directly obtained from the model, can be expressed as a sum over all of the photon emissions from excited levels *j* in all of the various charge states *q*+ of the ions in the plasma:

$$F = \sum_q \sum_{j,i} n_j^{q+} A_{ji} \qquad (4).$$

The sum in eq. (4) is carried out only over those transitions that produce photons in the measured 10 - 18 Å wavelength band. In the atomic-state model we actually calculate the relative level populations $n_j^{q+} / n^{q+}$, where $n^{q+}$ represents the total (summed over all of the levels of the ion) number density of $q$+ ions. This can be easily related to $F$ by writing:

$$n_j^{q+} = n_e \frac{n_H}{n_e} \frac{n_{Fe}}{n_H} \frac{n^{q+}}{n_{Fe}} \frac{n_j^{q+}}{n^{q+}} \qquad (5).$$

and hence,

$$F = n_e \frac{n_H}{n_e} \frac{n_{Fe}}{n_H} \sum_q \frac{n^{q+}}{n_{Fe}} \sum_{j,i} \frac{n_j^{q+}}{n^{q+}} A_{ji} \qquad (6).$$

The sum on the extreme right hand side of eq. (6), $\Sigma(n_j^{q+} / n^{q+})A_{ji}$ , is directly calculated in the model. The ionic fractional abundances $n^{q+} / n_{\mathrm{Fe}}$ are obtained, as described above, by fitting the calculated spectrum to the observed spectrum and by giving the appropriate weights to the various charge states. In the low-density approximation, the value of $n_e^2/F$, which is used for the *VEM* in eq. (3), is independent of the electron density assumed in the model. Standard solar values are used for the ratios $n_H / n_e$ and $n_{\mathrm{Fe}} / n_H$ . For a fully ionized plasma comprised of 90% hydrogen and 10% helium, $n_H / n_e = 0.83$. Following Feldman (1992), $n_{\mathrm{Fe}} / n_H$ is taken to be $3.2\times10^{-5}$. All of the values needed for calculating the *VEM* from the current observation, corresponding to the 10 – 18 Å wavelength range, are summarized in Table II. Finally, by using these quantities in eq. (3), one obtains a *VEM* ($n_e^2V$) of $10^{53}$ cm$^{-3}$ for the x-ray iron-L emitting regions of Capella. Considering the many different assumptions and estimated values involved in the calculation of the *VEM*, it is in reasonable agreement with the previously published total x-ray values for Capella of $9.5\times10^{52}$ cm$^{-3}$ (Mewe et al. 1982), $6.3\times10^{52}$ cm$^{-3}$ (Vedder & Canizares 1983), and $7.2\times10^{52}$ cm$^{-3}$ (Favata et al. 1997). The *VEM* values measured for individual EUV lines (Linsky et al. 1998) are slightly lower, about $4\times10^{52}$ cm$^{-3}$. The temperature dependent *VEM* distribution derived from the EUV line intensities (Brickhouse et al. 2000), peaks for ~550eV at a *VEM* value of ~$7\times10^{52}$ $cm^{-3}$. This value is also in very good agreement with our slightly higher result, since the additional contributions to the *VEM,* which Brickhouse et al. (2000) find at lower and higher temperature regions, in our model are all included at 600 eV. A *VEM* of $10^{53}$ $cm^{-3}$ and electron density of $10^{12}$ $cm^{-3}$ imply a volume of $10^{29}$ $cm^3$ for the iron-L emitting regions. This volume corresponds to a characteristic length ($V^{1/3}$) of ~$5\times10^9$ cm, which is much smaller than the average radius of the stars $R_* = 7.4\times10^{11}$cm (Hummell et al. 1994).

## 8. CONCLUSIONS

The x-ray line emission of Capella in the 10 – 18 Å wavelength range is analyzed by means of a simple coronal model for the $Fe^{15+}$ to $Fe^{21+}$ ions. The atomic data used for the Fe-L lines are found to be complete in accounting for all of the observed spectral features, including the weak ones. By employing the detailed level-by-level calculations, all of the iron lines in the 10 – 18 Å wavelength range can be identified. Furthermore, the calculated iron line intensities agree, in most cases, with the observed intensities to within the observational uncertainties (20%). It is shown that the observed iron spectrum of Capella can be reproduced by assuming one electron temperature and one electron density, although it is clear from observations in other wavelength bands that other temperature components are also present. The relatively large discrepancy of 30% between the observed intensity ratio of the strong line at 15.015 Å to the line at 15.264 Å and its calculated value is mostly attributed to inaccuracy in the calculated value and to blending with $Fe^{15+}$, but it also may be affected by self-absorption of the 15.015 Å line. Line absorption will soon be incorporated in our collisional-radiative model in a self-consistent manner, by including its influence on the level populations as well as on the escape of radiation from the plasma. The model can then be used to determine the importance of self-absorption processes on the Capella plasma as well as for other stellar coronae.

Using the intensities of lines corresponding to the 2s-3p and 2p-4d transitions in $Fe^{16+}$, an electron temperature of about 540 eV is measured. The comparison between the fractional ion abundances, which are derived from the spectrum, and those obtained from ionization balance calculations in the literature yields a temperature of about 600 eV. It is demonstrated that the iron lines in the wavelength range considered here do not show any sensitivity to the electron density and thus cannot be useful for density diagnostics. Finally, a volume emission measure of $10^{53}$ $cm^{-3}$ is calculated for the iron-L emitting regions of the Capella coronae.

We acknowledge generous support from the National Aeronautics and Space Administration. J.C. was funded by the NASA GSRP program under grant number NGT 5-50152. We thank David Huenemoerder for pointing out a technical error in an earlier version of this paper. We are grateful to an anonymous referee for helping us to significantly improve the presentation of this work.

## REFERENCES

Arnaud, M., & Raymond, J. 1992, ApJ, 398, 394

Bar-Shalom, A., Klapisch, M., & Oreg, J. 1988, Phys. Rev. A, 38, 1733

Bar-Shalom A., Klapisch M., Goldstein W.H., & Oreg J. 1998, *The HULLAC code for atomic physics*, unpublished.

Bauer, F., & Bregman, J.N. 1996, ApJ, 457, 382

Brickhouse, N.S., & Drake, J.J. 1999, in the proceedings of *Astrophysical Plasmas: Codes, Models & Observations*, ed. S.J. Arthur, N.S. Brickhouse, & J. Franco (Mexico City).

Brickhouse, N.S., Dupree, A.K., Edgar, R.J., Liedahl, D.A., Drake, S.A., White, N.E., & Singh, K.P. 2000, ApJ, 530, 387

Brinkman, A.C., et al. 2000, ApJ, 530, L111

Brown, G.V., Beiersdorfer, P., Liedahl, D.A., Widmann, K., & Kahn, S.M. 1998, ApJ, 502, 1015

Brown, G.V., Beiersdorfer, P., Liedahl, D.A., Widmann, K., & Kahn, S.M. 2000, LLNL preprint (UCRL-JC-136647)

Catura, R.C., Acton, L.W., & Johnson, H.M. 1975, ApJ, 196, L47

Chen, M.H., & Reed, K.J. 1986, Phys. Rev. A 40, 2292

Dupree, A.K., Brickhouse, N.S., Doschek, G.A., Green, J.C., & Raymond, J. 1993, ApJ, 418, L41

Favata, F., Mewe, R., Brickhouse, N.C., Pallavicini, R., Micela, G., & Dupree, A.K. 1997, A&A, 324, L37

Feldman, U., & Doschek, G.A. 1978, A&A 65, 215

Feldman, U., Doscheck, G.A., & Kreplin, R.W. 1982, ApJ, 260, 885

Feldman, U. 1992, Phys. Scr., 46, 202

Gu, M.F. et al. 1999, ApJ, 518, 1002

Hummel, C.A., Armstrong, J.T., Quirrenbach, A., Buscher, D.F., Mozurkewich, D., & Elias, N.M. 1994, Astron. J., 107, 1859

Klapisch, M. 1971, Comput. Phys. Commun. 2, 239

Klapisch, M., Schwob, J.L., Fraenkel, B., & Oreg, J. 1977, J. Opt. Soc. Am. 67, 148

Lemen, J.R., Mewe, R., Schrijver, C.J., & Fludra, A. 1989, ApJ, 341, 474

Linsky, J.L., Wood, B.E., Brown, A., & Osten, R.A. 1998, ApJ, 492, 767

Markert, T.H., Canizares, C.R., Dewey, D., McGuirk, M., Pak, C., & Schattenburg, M.L. 1995, Proc. SPIE, 2280, 168.

Mazzotta, P., Mazzitelli, G., Colafrancesco, S., & Vittorio, N. 1998, A&AS, 133, 403

McKenzie, D.L., Landecker, P.B., Broussard, R.M., Rugge, H.R., Young, R.M., Feldman, U., & Doschek, G.A. 1980, ApJ, 241, 409

Mewe, R., et al. 1982, ApJ, 260, 233

Raymond, J., & Smith, B.W. 1986, ApJ, 306, 762

Schrijver, C.J., Mewe, R., van den Oord, G.H.J, & Kaastra, J.S. 1995, A&A 302, 438

Schulz, N.S., Taylor, S.C., Dewey, D., & Marshall, H.L., available at http://space.mit.edu/HETG/xrcf.html

Vedder, P.W. & Canizares, C.R. 1983, ApJ, 270, 666

Waljeski, K., Moses, D., Dere, K.P., Saba, J.L.R, Web, D.F., & Zarro, D.M. 1994, ApJ, 429, 909

Table Captions:

Table I

The Capella spectral lines in the 10 - 18 Å wavelength range. Labels refer to the peaks of the observed spectrum in Figs. 1a and 1b. For each line, the observed wavelength $\lambda_{observed,}$ the solar wavelength $\lambda_{Sun}$ from McKenzie et al. (1980), the laboratory measured wavelength $\lambda_{EBIT}$ from Brown et al. (1998, 2000), and the presently calculated wavelength $\lambda_{HULLAC}$ are given. The upper and lower configurations of each transition are indicated together with the corresponding quantum numbers for the total angular momenta, $J_U$ and $J_L$. The number of counts in the (5mÅ) bin corresponding to the peak and the line flux obtained from the Gaussian fits are given in the last two columns. Strongly blended lines that do not have separate peaks are denoted by (b). Wide peaks with FWHM > 25 mÅ are denoted by (w).

Table II

Quantities used for calculating the volume emission measure of the emitting plasma. See definitions in text.

Figure Captions:

Figure 1:

Observed Capella spectrum (top) obtained by the HETG spectrometer using the MEG grating. Positive and negative first orders have been averaged. Labels correspond to the line identifications in Table I. The calculated iron spectrum is plotted at the bottom in arbitrary units, for comparison.

Figure 2:

The two-Gaussian fit to peak 12 (Fig. 1a) indicating the presence of two blended spectral lines 12a and 12b (see Table I).

Figure 3:

Observed Capella spectrum (top) compared to the total calculated spectrum. The seven separate plots at the bottom of the figure show the individual-ion contributions of $Fe^{15+}$ to $Fe^{21+}$. The spectra are calculated assuming plasma conditions of $kT_e$= 600 eV. The lines are given a uniform Gaussian shape with 15 mÅ FWHM. The non-iron lines and blends are indicated by the numbers 1-3, corresponding, respectively, to lines from Ne, Ni, and O.

Figure 4:

Observed Capella spectrum (top) compared to total spectra calculated for the same density, but for varying temperatures, $kT_e$= 200 eV, 600 eV, and 1000 eV. The ionic fractional abundances used for $Fe^{15+}$ through $Fe^{21+}$ are: 0.06, 0.34, 0.25, 0.13, 0.11, 0.06, and 0.05, respectively.

Figure 5:

Temperature sensitive intensity ratios of the $Fe^{16+}$ lines at 13.829 Å (2s-3p) and 12.266 Å (2p-4d) to the $Fe^{16+}$ line at 15.264 Å (2p-3d). The observed ratios, increased by 10% to account for the $Fe^{15+}$ contribution at 15.264 Å (see Sec. 5), are indicated by the horizontal lines. The calculated values are obtained by a coronal model including high Rydberg levels up to $n$= 12.

Figure 6:

Fractional abundances of iron ions $Fe^{15+}$ through $Fe^{21+}$. The squares (with 20% error bars) indicate the values derived from the Capella observation. The triangles, circles, and diamonds represent the calculated values of Mazzotta et al. (1998) for 430 eV, 610 eV, and 860 eV, respectively. Lines are drawn between the data points to guide the eye. The $Fe^{15+}$ abundance at 860 eV as well as the $Fe^{20+}$ and $Fe^{21+}$ abundances at 430 eV are too low to be seen.

Table I

| Label | $\lambda_{observed}$ (Å) | $\lambda_{Sun}$ (Å) | $\lambda_{EBIT}$ (Å) | $\lambda_{HULLAC}$ (Å) | Ion | Upper Configuration | $J_U$ | Lower Configuration | $J_L$ | Peak Counts | Flux ($10^{-4} s^{-1} cm^{-2}$) |
|---|---|---|---|---|---|---|---|---|---|---|---|
| 1 | 10.240 | 10.256 | | 10.238 | $Ne^{9+}$ | $3p_{3/2}$ | 3/2 | 1s | 1/2 | 34 | 1.32 |
| | | | | 10.239 | $Ne^{9+}$ | $3p_{1/2}$ | 1/2 | 1s | 1/2 | | |
| 2 | 10.768 | 10.778 | 10.770 | 10.782 | $Fe^{16+}$ | $2p_{1/2}{}^{2}2p_{3/2}{}^{3}6d_{5/2}$ | 1 | $2p_{1/2}{}^{2}2p_{3/2}{}^{4}$ | 0 | 22 | 1.08 $^{(w)}$ |
| 3 | 10.818 | 10.826 | 10.816 | 10.827 | $Fe^{18+}$ | $2p_{1/2}2p_{3/2}{}^{2}4d_{5/2}$ | 3 | $2p_{1/2}{}^{2}2p_{3/2}{}^{2}$ | 2 | 21 | 1.02 $^{(w)}$ |
| 4a | 10.997 | 10.990 | | 11.009 | $Ne^{8+}$ | $1s4p_{3/2}$ | 1 | $1s^2$ | 0 | 17 | 0.66 |
| 4b | 11.022 | 11.031 | 11.026 | 10.012 | $Fe^{16+}$ | $2s2p_{1/2}{}^{2}2p_{3/2}{}^{4}4p_{3/2}$ | 1 | $2s^{2}2p_{1/2}{}^{2}2p_{3/2}{}^{4}$ | 0 | 18 | 0.64 |
| 5 | 11.129 | 11.140 | 11.131 | 11.142 | $Fe^{16+}$ | $2p_{1/2}2p_{3/2}{}^{4}5d_{3/2}$ | 1 | $2p_{1/2}{}^{2}2p_{3/2}{}^{4}$ | 0 | 26 | 1.02 |
| 6 | 11.251 | 11.255 | 11.254 | 11.262 | $Fe^{16+}$ | $2p_{1/2}{}^{2}2p_{3/2}{}^{3}5d_{5/2}$ | 1 | $2p_{1/2}{}^{2}2p_{3/2}{}^{4}$ | 0 | 29 | 1.32 |
| 7 | 11.323 | 11.336 | 11.326 | 11.332 | $Fe^{17+}$ | $2p_{1/2}2p_{3/2}{}^{3}4d_{3/2}$ | 5/2 | $2p_{1/2}{}^{2}2p_{3/2}{}^{3}$ | 3/2 | 29 | 1.38 |
| | | | 11.326 | 11.336 | $Fe^{17+}$ | $2p_{1/2}2p_{3/2}{}^{3}4d_{5/2}$ | 3/2 | $2p_{1/2}{}^{2}2p_{3/2}{}^{3}$ | 3/2 | | |
| 8 | 11.422 | 11.445 | 11.423 | 11.435 | $Fe^{17+}$ | $2p_{1/2}2p_{3/2}{}^{3}4d_{3/2}$ | 5/2 | $2p_{1/2}{}^{2}2p_{3/2}{}^{3}$ | 3/2 | 23 | 1.02 |
| 9a | 11.527 | 11.537 | 11.527 | 11.537 | $Fe^{17+}$ | $2p_{1/2}{}^{2}2p_{3/2}{}^{2}4d_{5/2}$ | 5/2 | $2p_{1/2}{}^{2}2p_{3/2}{}^{3}$ | 3/2 | 38 | 1.49 |
| 9b | 11.549 | 11.537 | | 11.554 | $Ne^{8+}$ | $1s3p_{3/2}$ | 1 | $1s^2$ | 0 | 22 | 0.87 |
| 10a | 11.742 | 11.742 | 11.736 | 11.758 | $Fe^{22+}$ | $2s3d_{5/2}$ | 2 | $2s2p_{3/2}$ | 1 | 18 | 0.68 |
| 10b | 11.770 | 11.770 | 11.770 | 11.780 | $Fe^{21+}$ | $3d_{3/2}$ | 3/2 | $2p_{1/2}$ | 1/2 | 17 | 0.77 |
| 11a | 12.126 | 12.127 | 12.124 | 12.135 | $Fe^{16+}$ | $2p_{1/2}2p_{3/2}{}^4 4d_{3/2}$ | 1 | $2p_{1/2}{}^2 2p_{3/2}{}^4$ | 0 | 134 | 5.24 |
| 11b | 12.135 | 12.127 | | 12.132 | $Ne^{9+}$ | $2p_{3/2}$ | 3/2 | 1s | 1/2 | (b) | 4.97 |
| | | | | 12.137 | $Ne^{9+}$ | $2p_{1/2}$ | 1/2 | 1s | 1/2 | | |
| 12a | 12.266 | 12.276 | 12.266 | 12.272 | $Fe^{16+}$ | $2p_{1/2}{}^2 2p_{3/2}{}^3 4d_{5/2}$ | 1 | $2p_{1/2}{}^2 2p_{3/2}{}^4$ | 0 | 68 | 3.89 |
| 12b | 12.292 | 12.276 | 12.284 | 12.295 | $Fe^{20+}$ | $2p_{1/2}3d_{3/2}$ | 1 | $2p_{1/2}{}^2$ | 0 | 28 | 1.46 |
| 13a | 12.406 | 12.401 | | 12.399 | $Fe^{15+}$ | $2p_{1/2}2p_{3/2}{}^4 3s4d_{3/2}$ | 3/2 | $2p_{1/2}{}^2 2p_{3/2}{}^4 3s$ | 1/2 | 14 | 0.70 |
| 13b | 12.431 | 12.428 | | 12.431 | $Ni^{18+}$ | $2p_{1/2}2p_{3/2}{}^4 3d_{3/2}$ | 1 | $2p_{1/2}{}^2 2p_{3/2}{}^4$ | 0 | 37 | 2.17 |
| 14 | 12.650 | 12.657 | | 12.659 | $Ni^{18+}$ | $2p_{1/2}{}^2 2p_{3/2}{}^3 3d_{5/2}$ | 1 | $2p_{1/2}{}^2 2p_{3/2}{}^4$ | 0 | 13 | 0.99 (w) |
| 15a | 12.820 | 12.832 | 12.824 | 12.820 | $Fe^{19+}$ | $2p_{1/2}2p_{3/2}3d_{3/2}$ | 1/2 | $2p_{1/2}{}^2 2p_{3/2}$ | 3/2 | 32 | 2.65 (w) |
| 15b | 12.850 | 12.832 | 12.846 | 12.835 | $Fe^{19+}$ | $2p_{1/2}2p_{3/2}3d_{3/2}$ | 3/2 | $2p_{1/2}{}^2 2p_{3/2}$ | 3/2 | 26 | 2.42 (w) |
| | | 12.832 | 12.864 | 12.854 | $Fe^{19+}$ | $2p_{1/2}2p_{3/2}3d_{5/2}$ | 5/2 | $2p_{1/2}{}^2 2p_{3/2}$ | 3/2 | | |
| 16a | 13.446 | 13.450 | | 13.450 | $Ne^{8+}$ | $1s2p_{3/2}$ | 1 | $1s^2$ | 0 | 67 | 3.10 |
| 16b | 13.464 | 13.450 | 13.462 | 13.475 | $Fe^{18+}$ | $2p_{1/2}2p_{3/2}^{2}3d_{3/2}$ | 1 | $2p_{1/2}^{2}2p_{3/2}^{2}$ | 2 | 39 | 2.08 |
| | | | 13.462 | 13.475 | $Fe^{18+}$ | $2p_{1/2}2p_{3/2}^{2}3d_{3/2}$ | 1 | $2p_{1/2}^{2}2p_{3/2}^{2}$ | 0 | | |
| 17a | 13.508 | 13.517 | 13.497 | 13.516 | $Fe^{18+}$ | $2p_{1/2}2p_{3/2}^{2}3d_{3/2}$ | 2 | $2p_{1/2}^{2}2p_{3/2}^{2}$ | 2 | 56 | 2.07 |
| 17b | 13.524 | 13.517 | 13.518 | 13.525 | $Fe^{18+}$ | $2p_{1/2}^{2}2p_{3/2}3d_{5/2}$ | 3 | $2p_{1/2}^{2}2p_{3/2}^{2}$ | 2 | (b) | 2.97 |
| 18 | 13.550 | | | 13.569 | $Ne^{8+}$ | $1s2p_{1/2}$ | 1 | $1s^{2}$ | 0 | 22 | 2.57 (w) |
| 19 | 13.701 | 13.701 | | 13.715 | $Ne^{8+}$ | 1s2s | 1 | $1s^{2}$ | 0 | 31 | 4.90 (w) |
| 20a | 13.783 | 13.791 | | 13.793 | $Ni^{18+}$ | $2p_{1/2}2p_{3/2}^{4}3s$ | 1 | $2p_{1/2}^{2}2p_{3/2}^{4}$ | 0 | 26 | 2.21 |
| 20b | 13.797 | | 13.795 | 13.807 | $Fe^{18+}$ | $2p_{1/2}2p_{3/2}^{2}3d_{5/2}$ | 3 | $2p_{1/2}^{2}2p_{3/2}^{2}$ | 2 | 31 | 2.25 (w) |
| 21a | 13.829 | 13.847 | 13.825 | 13.793 | $Fe^{16+}$ | $2s2p_{1/2}^{2}2p_{3/2}^{4}3p_{3/2}$ | 1 | $2p_{1/2}^{2}2p_{3/2}^{4}$ | 0 | 27 | 3.12 |
| 21b | 13.854 | 13.847 | 13.839 | 13.860 | $Fe^{18+}$ | $2p_{1/2}^{2}2p_{3/2}3d_{5/2}$ | 2 | $2p_{1/2}^{2}2p_{3/2}^{2}$ | 2 | 22 | 0.74 |
| 22 | 14.041 | 14.037 | | 14.056 | $Ni^{18+}$ | $2p_{1/2}^{2}2p_{3/2}^{3}3s$ | 1 | $2p_{1/2}^{2}2p_{3/2}^{4}$ | 0 | 16 | 2.28 (w) |
| 23 | 14.080 | 14.082 | | 14.092 | $Ni^{18+}$ | $2p_{1/2}^{2}2p_{3/2}^{3}3s$ | 2 | $2p_{1/2}^{2}2p_{3/2}^{4}$ | 0 | 22 | 1.45 |
| 24 | 14.206 | 14.220 | 14.208 | 14.205 | $Fe^{17+}$ | $2p_{1/2}2p_{3/2}^{3}3d_{3/2}$ | 5/2 | $2p_{1/2}^{2}2p_{3/2}^{3}$ | 3/2 | 116 | 11.10 |
| | | | 14.208 | 14.218 | $Fe^{17+}$ | $2p_{1/2}2p_{3/2}^{3}3d_{5/2}$ | 3/2 | $2p_{1/2}^{2}2p_{3/2}^{3}$ | 3/2 | | |
| 25 | 14.261 | 14.281 | 14.256 | 14.276 | $Fe^{17+}$ | $2p_{1/2}2p_{3/2}^{3}3d_{3/2}$ | 1/2 | $2p_{1/2}^{2}2p_{3/2}^{3}$ | 3/2 | 39 | 3.47 |

| | | | | | | | | | | |
|---|---|---|---|---|---|---|---|---|---|---|
| | | | 14.267 | 14.286 | $Fe^{19+}$ | $2s2p_{1/2}^{2}2p_{3/2}3s$ | 3/2 | $2s2p_{1/2}^{2}2p_{3/2}^{2}$ | 5/2 | | |
| 26 | 14.374 | 14.388 | 14.373 | 14.387 | $Fe^{17+}$ | $2p_{1/2}2p_{3/2}^{3}3d_{3/2}$ | 5/2 | $2p_{1/2}^{2}2p_{3/2}^{3}$ | 3/2 | 42 | 3.85 |
| 27a | 14.535 | 14.549 | 14.534 | 14.549 | $Fe^{17+}$ | $2p_{1/2}^{2}2p_{3/2}^{2}3d_{5/2}$ | 5/2 | $2p_{1/2}^{2}2p_{3/2}^{3}$ | 3/2 | 20 | 2.13 |
| 27b | 14.552 | | 14.571 | 14.569 | $Fe^{17+}$ | $2p_{1/2}^{2}2p_{3/2}^{2}3d_{5/2}$ | 3/2 | $2p_{1/2}^{2}2p_{3/2}^{3}$ | 3/2 | 19 | 1.53 |
| 28 | 14.669 | 14.671 | 14.664 | 14.681 | $Fe^{18+}$ | $2p_{1/2}2p_{3/2}^{2}3s$ | 3 | $2p_{1/2}^{2}2p_{3/2}^{2}$ | 2 | 14 | 1.29 |
| 29 | 15.015 | 15.013 | 15.014 | 15.014 | $Fe^{16+}$ | $2p_{1/2}2p_{3/2}^{4}3d_{3/2}$ | 1 | $2p_{1/2}^{2}2p_{3/2}^{4}$ | 0 | 291 | 30.0 |
| 30 | 15.082 | 15.083 | 15.079 | 15.110 | $Fe^{18+}$ | $2p_{1/2}^{2}2p_{3/2}3s$ | 2 | $2p_{1/2}^{2}2p_{3/2}^{2}$ | 2 | 39 | 3.60 |
| 31 | 15.181 | | | 15.175 | $O^{7+}$ | $4p_{3/2}$ | 3/2 | 1s | 1/2 | 24 | 1.98 |
| | | | | 15.176 | $O^{7+}$ | $4p_{1/2}$ | 1/2 | 1s | 1/2 | | |
| | | | | 15.188 | $Fe^{15+}$ | $2p_{1/2}2p_{3/2}^{4}3s3d_{3/2}$ | 1/2 | $2p_{1/2}^{2}2p_{3/2}^{4}3s$ | 1/2 | | |
| 32 | 15.207 | 15.198 | 15.198 | 15.225 | $Fe^{18+}$ | $2s2p_{1/2}^{2}2p_{3/2}^{2}3s$ | 2 | $2s2p_{1/2}^{2}2p_{3/2}^{3}$ | 2 | 26 | 2.29 |
| 33 | 15.264 | 15.262 | | 15.250 | $Fe^{15+}$ | $2p_{1/2}^{2}2p_{3/2}^{3}3s3p_{3/2}$ | 3/2 | $2p_{1/2}^{2}2p_{3/2}^{4}3s$ | 1/2 | 117 | 12.4 |
| | | | 15.261 | 15.272 | $Fe^{16+}$ | $2p_{1/2}^{2}2p_{3/2}^{3}3d_{5/2}$ | 1 | $2p_{1/2}^{2}2p_{3/2}^{4}$ | 0 | | |
| 34 | 15.454 | 15.456 | 15.453 | 15.470 | $Fe^{16+}$ | $2p_{1/2}^{2}2p_{3/2}^{3}3d_{3/2}$ | 1 | $2p_{1/2}^{2}2p_{3/2}^{4}$ | 0 | 26 | 2.70 |
| 35 | 15.628 | 15.630 | 15.625 | 15.642 | $Fe^{17+}$ | $2p_{1/2}2p_{3/2}^{3}3s$ | 5/2 | $2p_{1/2}^{2}2p_{3/2}^{3}$ | 3/2 | 46 | 4.03 |
| 36 | 15.827 | 15.827 | 15.824 | 15.854 | $Fe^{17+}$ | $2p_{1/2}2p_{3/2}{}^{3}3s$ | 3/2 | $2p_{1/2}{}^{2}2p_{3/2}{}^{3}$ | 3/2 | 27 | 2.97 |
| 37 | 15.875 | 15.869 | 15.870 | 15.884 | $Fe^{17+}$ | $2p_{1/2}2p_{3/2}{}^{3}3s$ | 3/2 | $2p_{1/2}2p_{3/2}{}^{4}$ | 1/2 | 28 | 3.41 |
| 38 | 16.008 | 16.004 | | 16.005 | $O^{7+}$ | $3p_{3/2}$ | 3/2 | 1s | 1/2 | 67 | 9.05 |
| | | | | 16.006 | $O^{7+}$ | $3p_{1/2}$ | 1/2 | 1s | 1/2 | | |
| | | | 16.004 | 16.024 | $Fe^{17+}$ | $2p_{1/2}{}^{2}2p_{3/2}{}^{2}3s$ | 3/2 | $2p_{1/2}{}^{2}2p_{3/2}{}^{3}$ | 3/2 | | |
| 39a | 16.078 | 16.070 | 16.071 | 16.098 | $Fe^{17+}$ | $2p_{1/2}{}^{2}2p_{3/2}{}^{2}3s$ | 5/2 | $2p_{1/2}{}^{2}2p_{3/2}{}^{3}$ | 3/2 | 76 | 9.60 |
| 39b | 16.111 | 16.108 | 16.110 | 16.132 | $Fe^{18+}$ | $2s^{2}2p_{1/2}2p_{3/2}{}^{2}3p_{1/2}$ | 2 | $2s2p_{1/2}{}^{2}2p_{3/2}{}^{3}$ | 2 | 20 | 2.32 |
| 40 | 16.170 | 16.165 | 16.159 | 16.198 | $Fe^{17+}$ | $2s2p_{1/2}2p_{3/2}{}^{4}3s$ | 3/2 | $2s2p_{1/2}{}^{2}2p_{3/2}{}^{4}$ | 1/2 | 16 | 2.23 (w) |
| 41 | 16.779 | 16.775 | 16.780 | 16.796 | $Fe^{16+}$ | $2p_{1/2}2p_{3/2}{}^{4}3s$ | 1 | $2p_{1/2}{}^{2}2p_{3/2}{}^{4}$ | 0 | 141 | 20.9 |
| 42 | 17.054 | 17.043 | 17.051 | 17.071 | $Fe^{16+}$ | $2p_{1/2}{}^{2}2p_{3/2}{}^{3}3s$ | 1 | $2p_{1/2}{}^{2}2p_{3/2}{}^{4}$ | 0 | 193 | 25.2 |
| 43 | 17.099 | 17.084 | 17.096 | 17.118 | $Fe^{16+}$ | $2p_{1/2}{}^{2}2p_{3/2}{}^{3}3s$ | 2 | $2p_{1/2}{}^{2}2p_{3/2}{}^{4}$ | 0 | 157 | 24.3 |
| 44 | 17.625 | 17.617 | 17.623 | 17.674 | $Fe^{17+}$ | $2s^{2}2p_{1/2}2p_{3/2}{}^{3}3p_{1/2}$ | 3/2 | $2s2p_{1/2}{}^{2}2p_{3/2}{}^{4}$ | 1/2 | 21 | 3.82 |

Table II

| Quantity | Value | Obtained from: |
|---|---|---|
| $F$ | 0.037 (photons) $sec^{-1}cm^{-2}$ | Observation |
| $\sum_q \frac{n^{q+}}{n_{Fe}} \sum_j \frac{n_j^{q+}}{n^{q+}} A_{ji}$ | 2.62 (photons) $sec^{-1}$ | The model (second sum) and fitting to observation |
| $n_H / n_e$ | 0.83 | Solar standard |
| $n_{Fe} / n_H$ | $3.2\times10^{-5}$ | Feldman 1992, Solar value |
| $F$ | $6.96\times10^{5}$ (photons)$sec^{-1}cm^{-3}$ | Eq. (6) |
| $d$ | 13.3pc= $4.1\times10^{19}$cm | Hummel et al. 1994 |

Behar et al. Fig. 1a

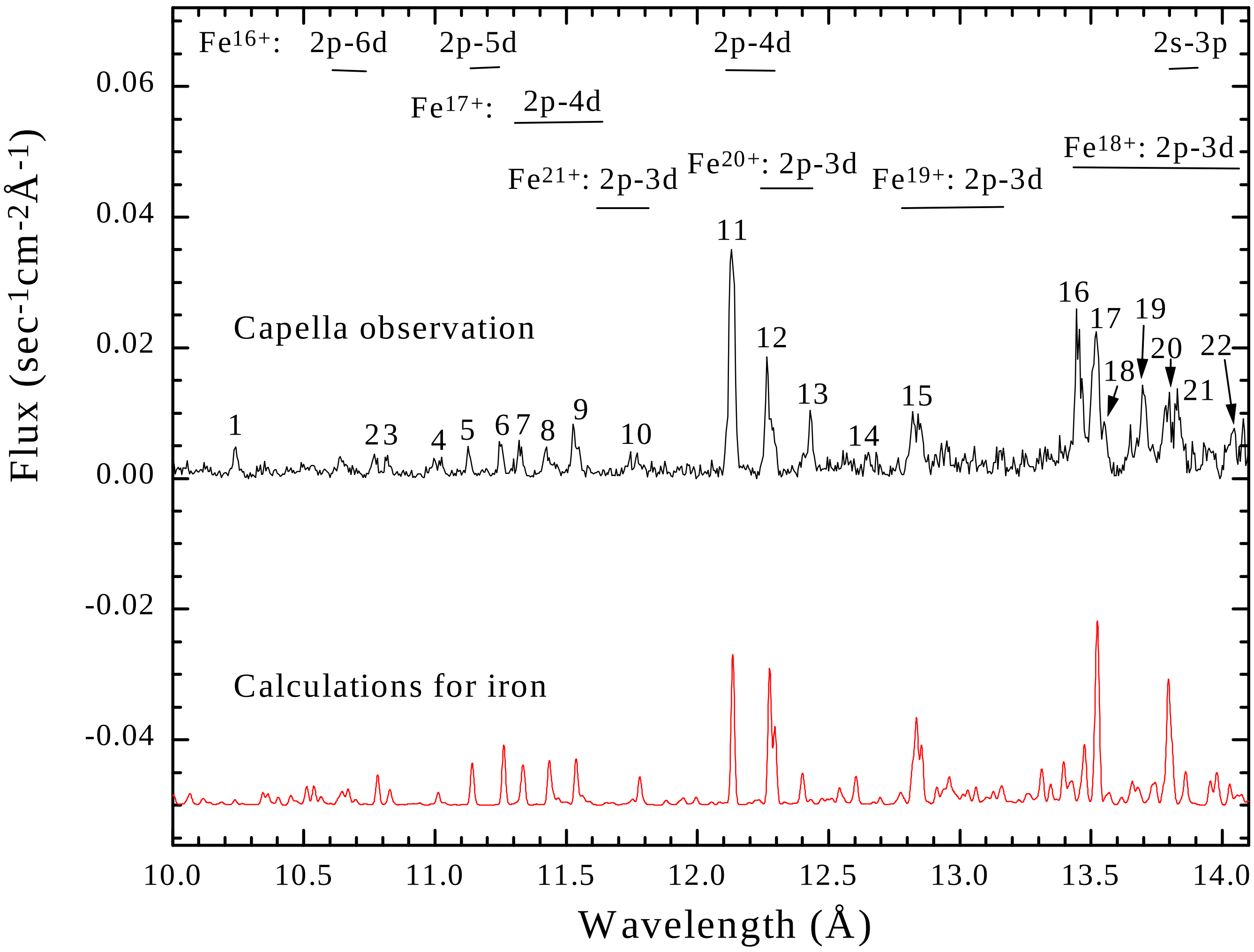

Behar et al. Fig. 1b

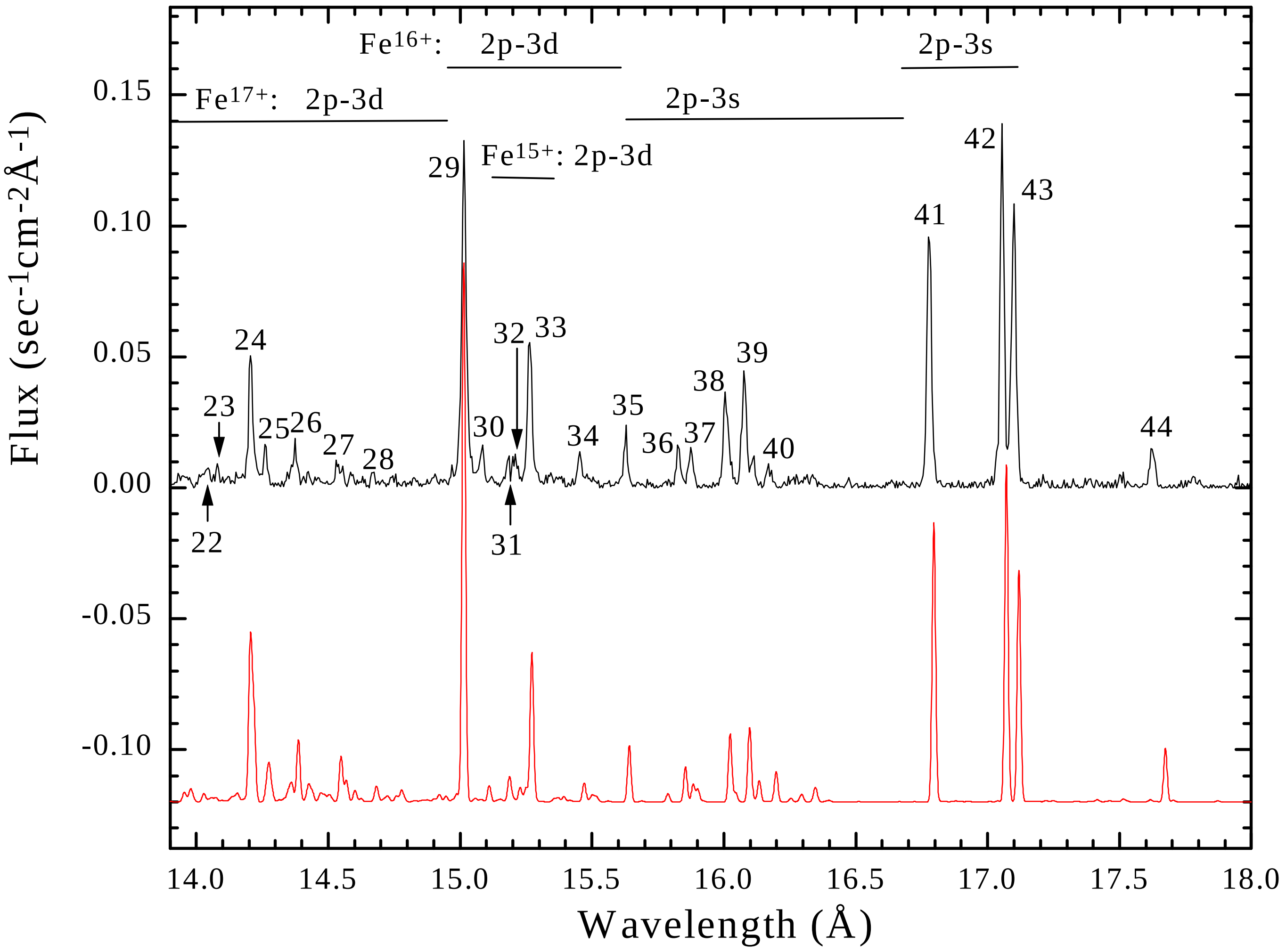

Behar et al. Fig. 2

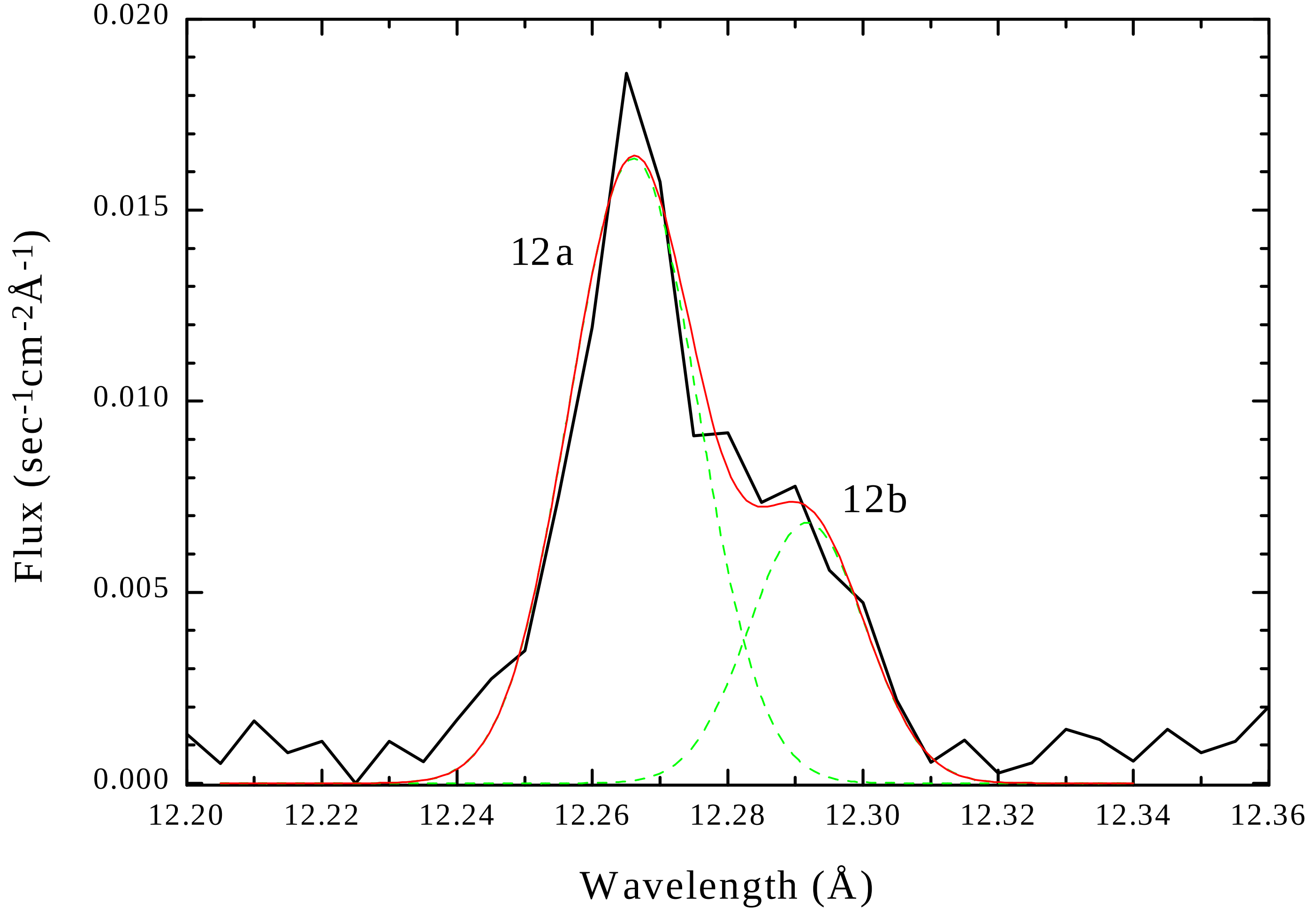

Behar et al. Fig. 3

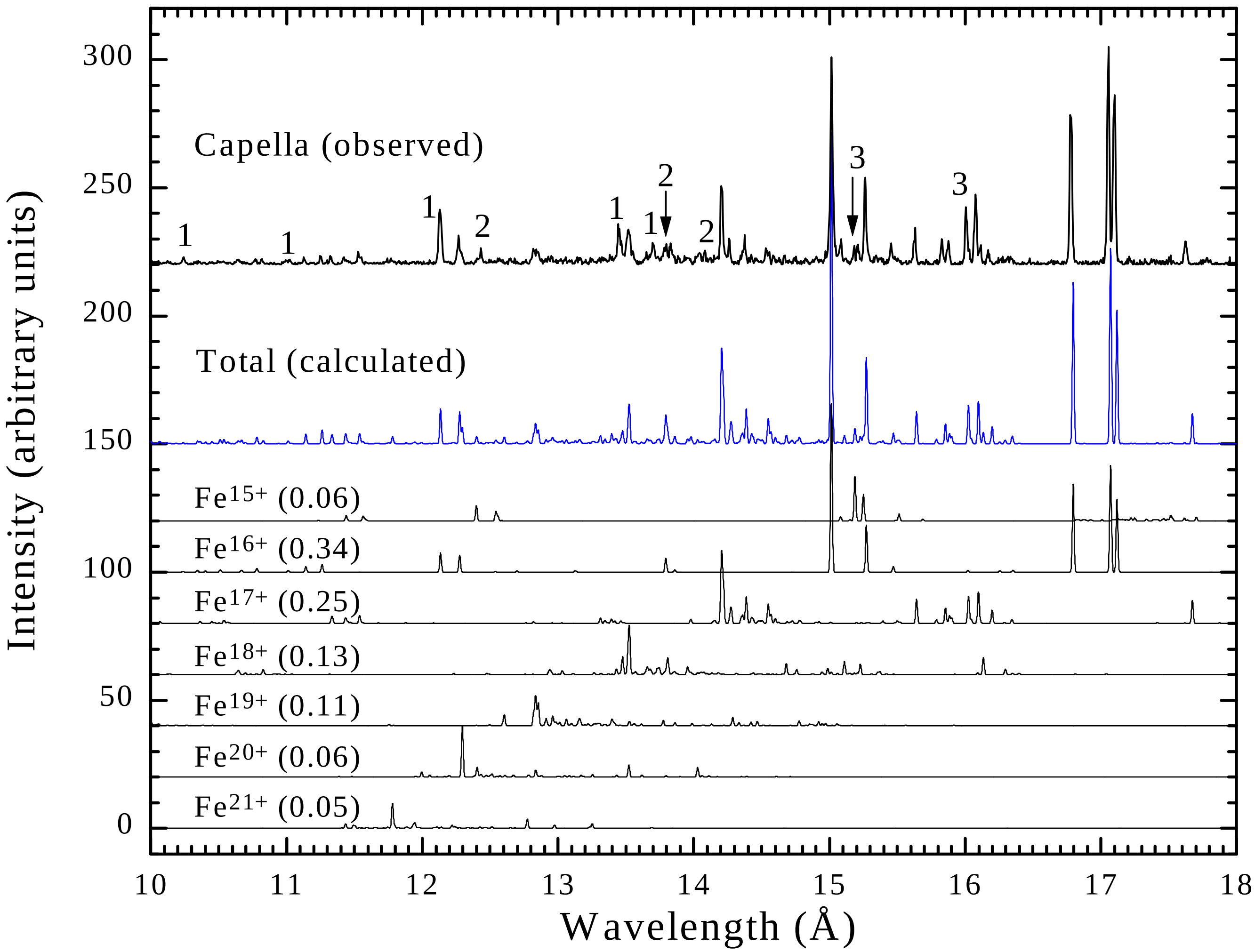

Behar et al. Fig. 4

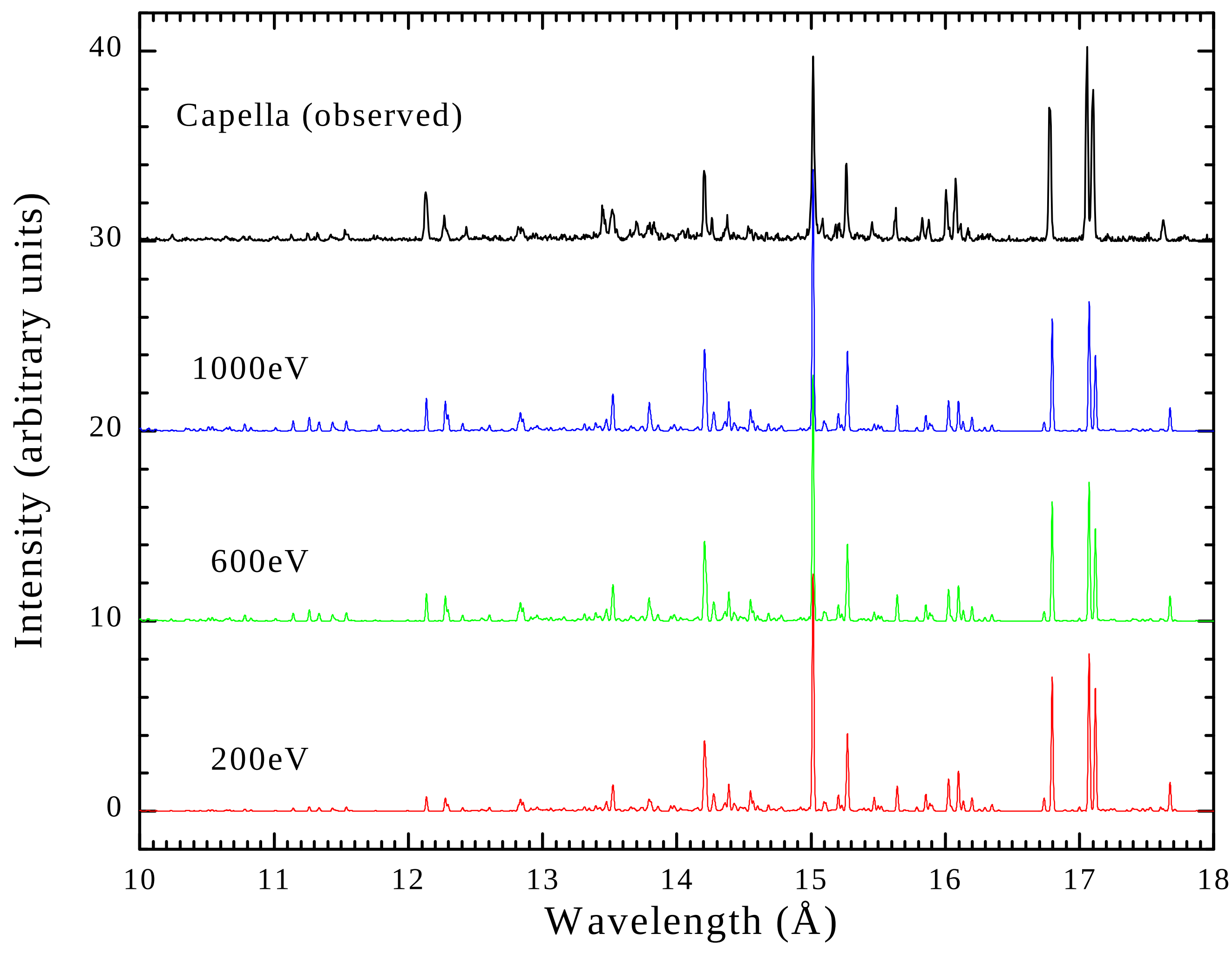

Behar et al. Fig. 5

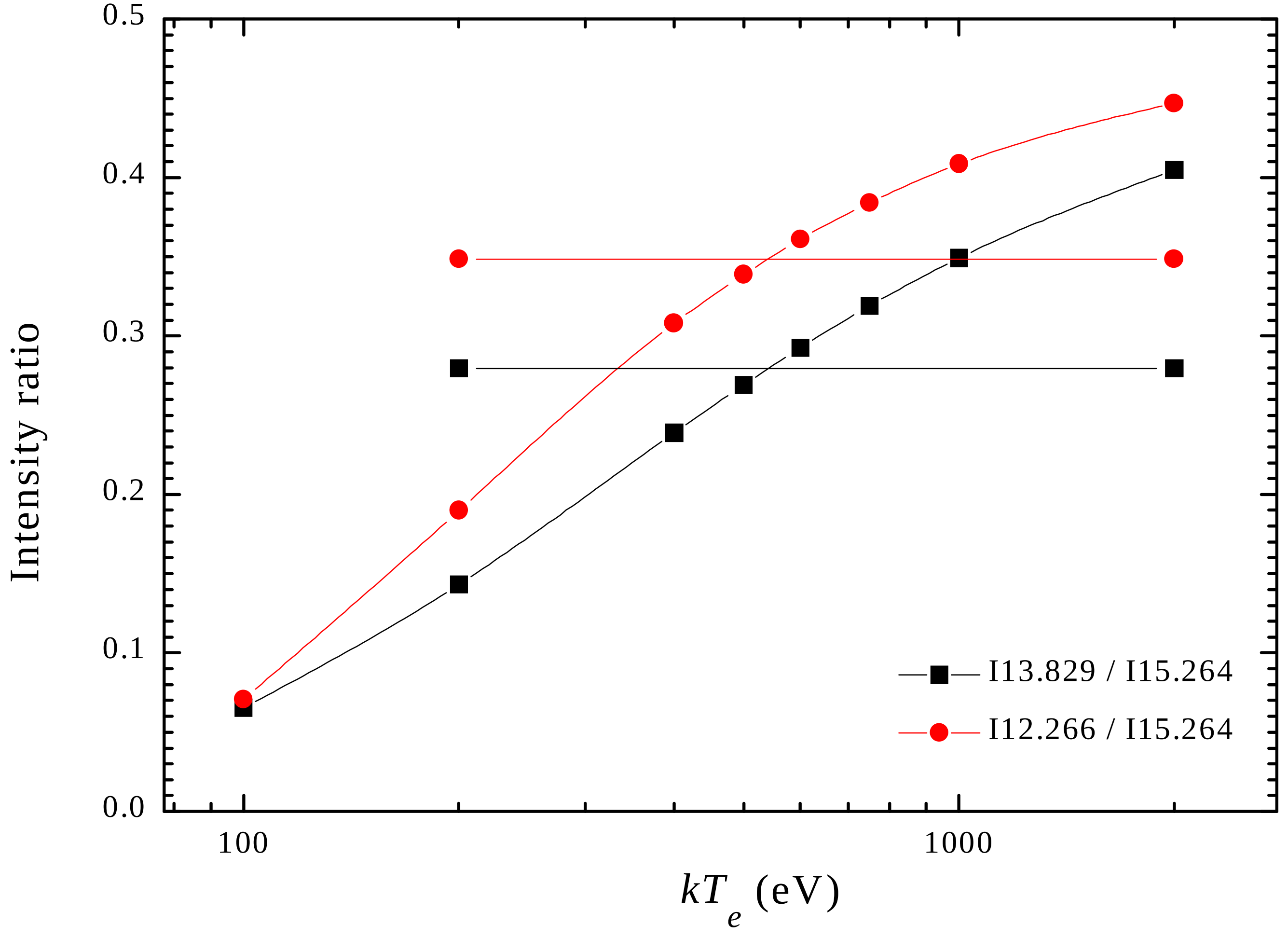

Behar et al. Fig. 6

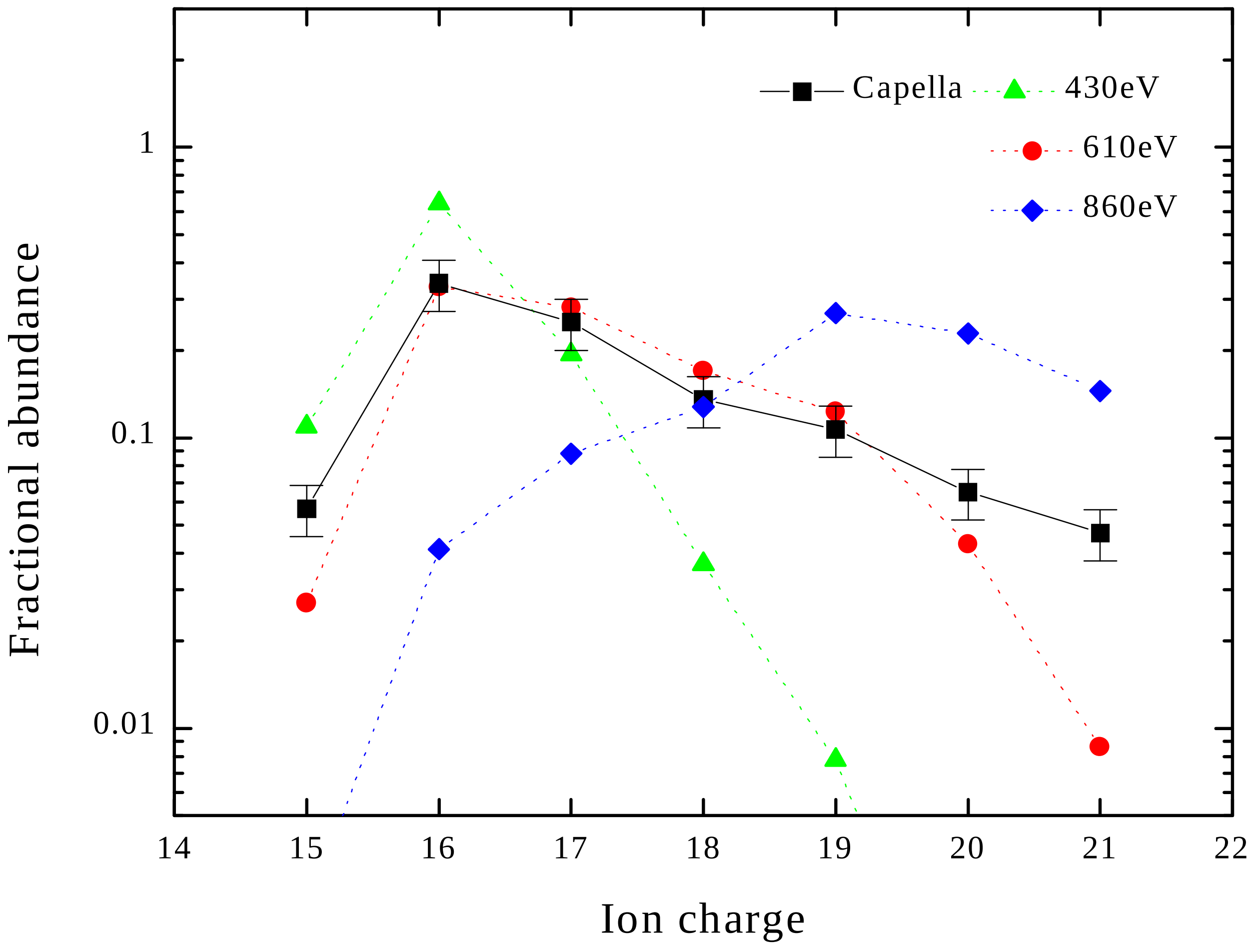